\documentclass{elsarticle}

\usepackage{graphicx} 
\usepackage{booktabs}
\usepackage{amsmath}

\title{
Flow transition of liquid–liquid two-phase Taylor–Couette flow with axial flow
}

\author[aff1]{Hayato Masuda\corref{cor1}}
\ead{hayato-masuda@omu.ac.jp}

\author[aff1]{Yuta Kikuchi}
\author[aff1]{Hiroyuki Iyota}

\address[aff1]{Department of Mechanical Engineering, Graduate School of Engineering, Osaka Metropolitan University, 1-1 Gakuen-cho, Naka-ku, Sakai-shi, Osaka 599-8531, Japan}

\cortext[cor1]{Corresponding author: hayato-masuda@omu.ac.jp}

\begin{document}

\begin{abstract}
A liquid–liquid two-phase Taylor–Couette flow is of fundamental and practical importance in multiphase flow systems. This study investigates the flow transitions in such a system with constant axial flow and gradually increased rotation of the inner cylinder. To clarify the effect of continuous-phase viscosity on flow transitions, several concentrations of glycerol–water solutions were employed as the continuous phase, while soybean oil was used as the dispersed phase. Visualization experiments revealed that the flow undergoes a cascade of transitions: stratified flow, disturbed stratified flow, transitional droplet flow (observed only in limited cases), unstable banded flow, and stable banded flow. A flow map was constructed using two dimensionless parameters, Reynolds number (\textit{Re}) and Weber number (\textit{We}). The transitions from disturbed stratified to unstable banded flow (TR I) and from unstable to stable banded flow (TR II) were found to be approximately expressed by a power-law relation of the form $\textit{We}\sim \textit{Re}^{\textit{A}}$, where \textit{A} is a fitting exponent. For TR I, A showed no clear dependence on continuous-phase viscosity, suggesting the combined effects of inertia, viscosity, and interfacial tension. In contrast, A decreased with increasing viscosity in TR II, becoming nearly zero at high glycerol concentration. Furthermore, to clarify the mechanism of stable banded flow formation, the droplet inertia was evaluated using the Stokes number (\textit{St}). It was revealed that the stable banded structure results from the loss of droplet inertia, associated with droplet micronization under higher rotation and the increasing inertia of the continuous phase.
\end{abstract}


\date{\empty}

\maketitle

\section{Introduction}
A Taylor–Couette (TC) flow caused between coaxial cylinders with the inner one rotating is one of the classical fluid mechanics problems. So far, numerous works on single-phase TC flow have been extensively carried out, and these reveal significant topics, e.g., flow transition (Dandelia et al., 2022), multistability (Wen et al., 2020), and heat/mass transfer (Masuda et al., 2019a); however, for the multiphase TC flow, even fundamental issues such as flow transition is still far from being completely understood despite a lot of published papers. Too many factors affect the multiphase TC flow dynamics, e.g., system (gas-liquid, liquid-liquid, solid-liquid, etc.), species of continuous and disperse phases, and differences in physical properties between them. In addition to scientific interest, the multiphase TC flow is also a hot topic for engineers because it has the potential to be a highly efficient apparatus owing to (i) the effective (but gentle) mixing characteristics, (ii) the excellent heat/mass transfer at the outer cylinder surface, and (iii) the ideal plug-flow properties when axial flow is imposed. Indeed, the TC flow has been applied to multiphase flow processes, e.g., reactors (Masuda et al., 2013; Matsumoto et al., 2021), contactors (Hubacz and Wroński, 2004; Dluska and Markowska-Radomsk, 2010), sterilizers (Masuda et al., 2019b), and particle processors (Wang and Tao, 2022; Zhang et al., 2024). Note that the TC flow with axial flow is often called the Taylor–Couette–Poiseuille (TCP) flow.

In practical terms, liquid–liquid two-phase TCP flow is one of the topics that should be prioritized in multiphase TCP flow systems due to its applicability to various processes. Davis and Weber (1960) reported a pioneering work. They studied the liquid-liquid two-phase TC flow dynamics of the nitric acid–tributyl phosphate system and showed high extraction performance at a high rotational speed of the inner cylinder. Following their work, Joseph and his co-workers theoretically and experimentally investigated the instability of the liquid-liquid interface at a relatively low rotational speed of the inner cylinder, that is the two fluids are stratified (Joseph et al., 1984; Renardy and Joseph, 1985; Joseph et al., 1985). They defined a non-dimensional number, the so-called Joseph factor, and derived a critical it at which the interface begins to be disturbed. Their works urge attempts to apply the two-phase liquid-liquid TC flow as the contactor (e.g., Baier et al., 2000; Sathe et al., 2010; Aksamija et al., 2015; Grafschafter et al., 2016). In that case, the larger specific interfacial area is preferable to enhance mass transfer between continuous and dispersed phases. According to Vigil and his co-workers, the dispersed phase is formed to shape into significantly fine droplets at a relatively higher rotational speed of the inner cylinder. For example, Campero and Vigil (1997) reported patterns of the TCP flow in the kerosene–water system at a relatively high rotational speed of the inner cylinder. They found, above a certain rotational speed, a unique flow pattern in which aqueous and organic-rich vortices alternate. They named the flow pattern a “banded” flow. In the banded flow, the dispersed phase is sufficiently micronized in the shape of droplets (Zhu and Vigil, 2001). Due to attaining the large specific interfacial area, this flow regime (i.e., banded flow) is considered useful for utilizing as the TCP contactor. Therefore, clarifying the conditions at which the banded flow is formed is a crucial issue for the liquid-liquid two-phase TCP flow problem.

In the early work by Campero and Vigil (1997), the flow diagram was reported based on the dispersed phase fraction (kerosene in their work) and the rotational speed of the inner cylinder. The diagram provides the approximate conditions to attain the banded flow. Thereafter, they extended the diagram by varying the dispersed phase species to clarify the effect of their physical properties on the flow pattern (Campero and Vigil, 1999). The diagram pattern was similar to the previous one, while the transition conditions to the banded flow slightly depended on the dispersed phase species. Additionally, the detailed mechanism for the banded structure formation was analyzed with the support of CFD simulation (Zhu and Vigil, 2001). Their work showed that the banded structure resulted from the migration of less dense dispersed phase droplets to Taylor vortex core due to the centrifugal forces.

Based on worthwhile works 30 or 40 years ago, the liquid–liquid two-phase TC flow is still being intensively studied (Kusumastuti et al., 2019; Campbell et al., 2019; Campbell et al., 2020; Hori et al., 2023; Li et al., 2023; Zeng et al., 2024). Nevertheless, the effect of the physical properties of dispersed/continuous phases, which Campero and Vigil (1997) pointed out as the future remark for the practical application, on the flow pattern remains unclear. Therefore, this study aimed to clarify the effect of continuous phase property on the liquid–liquid TCP flow dynamics using glycerol aqueous solutions with various concentrations. By adjusting the glycerol concentration, the density and viscosity can be varied in a wide range. In addition, the surface tension was also adjusted by varying the surfactant concentration. Among fundamental issues, this study examined the dynamical behavior of the TCP flow pattern with a gradual acceleration in the inner cylinder rotation based on visualization experiments.

\section{Experiment}
The experimental setup is depicted in Fig. 1. The TC flow apparatus consisted of a rotating inner cylinder with a radius of $R_\mathrm{i} = 0.0125\ \mathrm{m}$ and a stationary outer cylinder with a radius of $R_\mathrm{o} = 0.0175\ \mathrm{m}$, resulting in a gap width of $\textit{d} = 0.005\ \mathrm{m}$. The length of both cylinders, \textit{L}, was 0.3\ m. The rotational speed of the inner cylinder, $\omega$, was stepwisely increased from 3.11\ rad/s to 108.15\ rad/s every 3.0\ rad/s with an interval of 20\ s. A glycerol aqueous solution (continuous phase) was fed from the inlet to the apparatus at $Q_\mathrm{c} = 50\ \mathrm{mL/min}$. The four kinds of glycerol concentration (${C}_\mathrm{g}$) were used:$C_\mathrm{g} = 0, 20, 40, 60\mathrm{wt\%}$. To vary the surface tension, sodium dodecyl sulfate with various concentrations, ${C}_\mathrm{s}$, was added to the glycerol aqueous solution. The concentration is expressed in a non-dimensional form, dividing by a critical micelle concentration, ${C}_\mathrm{cr}$. ${C}_\mathrm{s}/{C}_\mathrm{cr}$ was adjusted from 0.2\ to\ 1.2. In the central part of experiments, a soybean oil (dispersed phase) was introduced to the annular space at ${Q}_\mathrm{d} = 5\ \mathrm{mL/min}$ from a nozzle 0.04\ m apart from the inlet, that is, a volumetric flow rate ratio, $\varepsilon = {Q}_\mathrm{d} / {Q}_\mathrm{c}$, was\ 0.1. Only for the measurement of droplet sizes in the flow regime in which the dispersed phase is formed to shape into droplets, ${Q}_\mathrm{d}$ was reduced to ${Q}_\mathrm {d} = 2.5\ \mathrm{ml/min}$ (i.e., $\varepsilon = 0.05$) to improve ease of detection and measurement for each droplet. The continuous and dispersed phase fluids were dyed with uranin and Oil Red O to distinguish each phase. The flow pattern was recorded using a high-resolution video camera (HC-VX992M, Panasonic Co.).

\begin{figure}[htbp]
  \centering
  \includegraphics[width=1\linewidth]{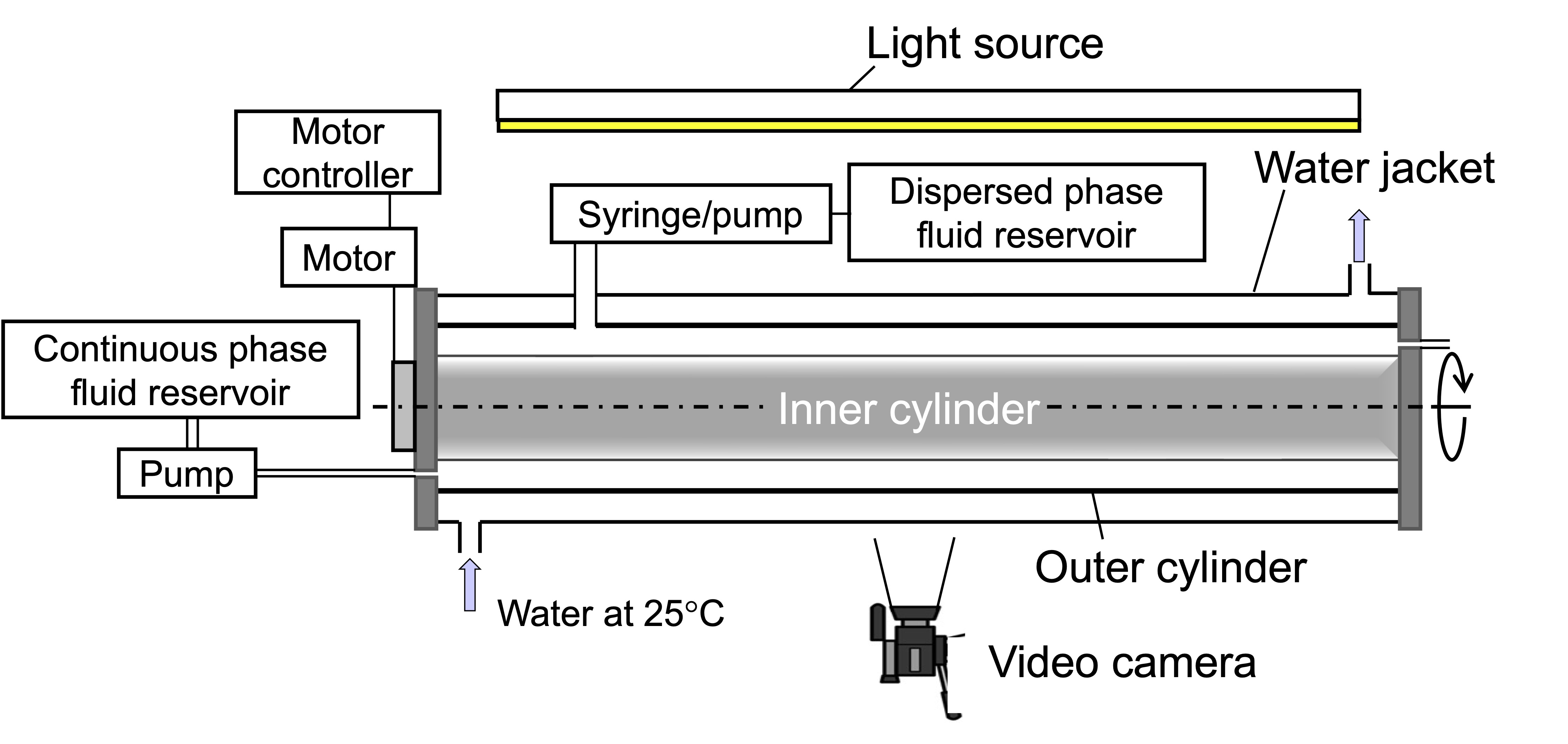}
  \caption{Experimental apparatus.}
  \label{fig:setup}
\end{figure}

Physical properties of each fluid, density, viscosity, and interfacial tension, were measured at room temperature using a density meter (DA-640, Kyoto Electronics Manufacturing Co., Ltd.), a rheometer (MCR102, Anton Paar GmbH), and a contact angle meter (B100, ASUMI GIKEN Ltd.), respectively. The density and viscosity of fluids and the interfacial tension of soybean oil (dispersed phase) with the various continuous-phase fluids were tabulated in Tables 1\ and 2.

\begin{table}[htbp]
  \centering
  \caption{Physical properties of glycerol aqueous solution and soybean oil.}
  \label{tab:properties}
  \begin{tabular}{l c c c}
    \toprule
    Fluid & $C_g$ [wt\%] & Density, $\rho$ [kg/m$^3$] & Viscosity, $\mu$ [Pa$\cdot$s] \\
    \midrule
    Glycerol aqueous solution & 0  & 998.0  & $9.98 \times 10^{-4}$ \\
                              & 20 & 1045.5 & $1.59 \times 10^{-3}$ \\
                              & 40 & 1094.5 & $3.80 \times 10^{-3}$ \\
                              & 60 & 1150.9 & $1.08 \times 10^{-2}$ \\
    \midrule
    Soybean oil & -- & 920.0 & $5.22 \times 10^{-2}$ \\
    \bottomrule
  \end{tabular}
\end{table}

\begin{table}[htbp]
  \centering
  \caption{Interfacial tension of soybean oil in various continuous-phase fluids.}
  \label{tab:interfacial_tension}
  \begin{tabular}{c c c c c}
    \toprule
    $C_g$ [wt\%] & \multicolumn{4}{c}{Interfacial tension, $\sigma$ [mN/m]} \\
    \cmidrule(lr){2-5}
     & $C_s/C_{cr}=0$ & 0.4 & 0.8 & 1.2 \\
    \midrule
    0  & 26.24 & 14.02 & 6.98 & 2.56 \\
    20 & 19.50 & 10.00 & 3.55 & 2.33 \\
    40 & 18.15 & 10.06 & 6.36 & 3.70 \\
    60 & 17.22 & 10.52 & 8.02 & 5.86 \\
    \bottomrule
  \end{tabular}
\end{table}

\section{Results and discussion}
The liquid-liquid two-phase TC flow exhibited a cascade-type flow transition as the angular velocity of the inner cylinder $\omega$ increased. The flow patterns were broadly classified into stratified flow at lower $\omega$ and banded flow at higher $\omega$. As a representative example, the sequence of flow transitions for $C_\mathrm{g} = 40\mathrm{wt\%}$ and $C_\mathrm{s} / C_\mathrm{cr} = 1.2$ are shown in Fig. 2. At $\omega$ = 3.11 rad/s (Fig. 2 (a)), both fluids flowed in a stratified pattern maintaining the stable interface. The interface was destabilized by increasing $\omega$. At $\omega$ = 18.66 rad/s (Fig. 2 (b)), the stronger centrifugal force disrupted the interface, breaking it into several segments. These flow patterns are referred to as “stratified flow” and “disturbed stratified flow”, respectively, in this study. At a slightly higher angular velocity ($\omega$= 24.88 rad/s), droplets began to form from the broken interface segments (Fig. 2 (c)). This intermediate regime is defined as “transitional droplet flow”, as it represented a short-lived state in which discrete droplets begin to emerge from the disrupted stratified interface. This regime occurred only under limited conditions (i.e., specific conditions of $C_\mathrm{g}$ and $C_\mathrm{s} / C_\mathrm{cr}$), and it typically preceded the formation of unstable banded structures. As $\omega$ further increased, the generated droplets gradually organized into a banded structure, similar to the pattern reported by Vigil’s group (Campero and Vigil, 1999; Zhu and Vigil, 2001). However, at $\omega$ = 40.43 rad/s (Fig. 2 (d)), although band-like distributions were present, the droplet sizes were non-uniform, and the band boundaries were unclear due to irregular droplet motion. This regime is termed “unstable banded flow”. With a further increase in $\omega$, the banded structures became more distinct and stable, as shown in Fig. 2 (e) at $\omega$ = 77.75 rad/s. This regime is termed “stable banded flow”. Thus, the flow patterns observed in this study were classified into five regimes.

\begin{figure}[htbp]
  \centering
  \includegraphics[width=1\linewidth]{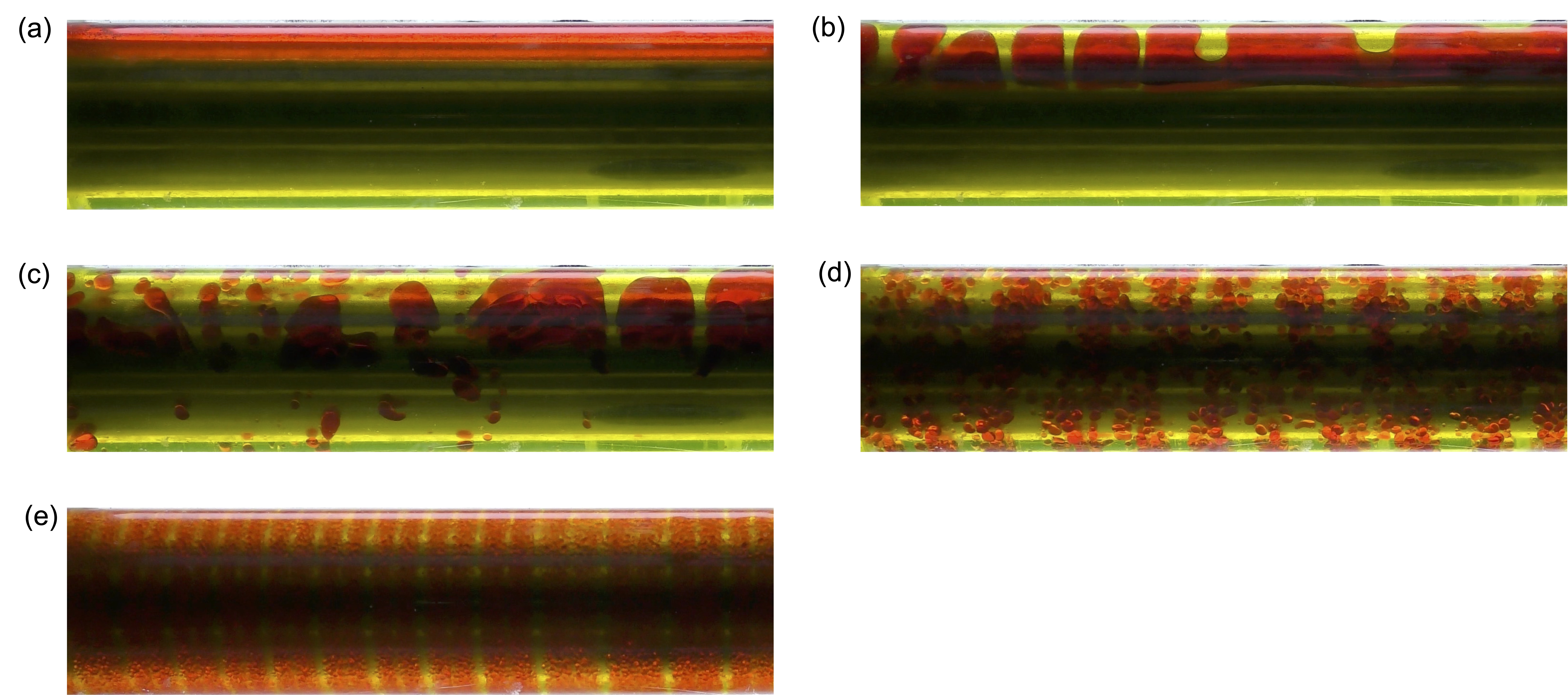}
  \caption{Flow pattern at $C_\mathrm{g}$ = 40wt\% and $C_\mathrm{s} / C_\mathrm{cr}$ = 1.2 : (a) stratified flow ($\omega$ = 3.11 rad/s), (b) disturbed stratified flow ($\omega$ = 18.66 rad/s), (c) transitional droplet flow ($\omega$ = 24.88 rad/s), (d) unstable banded flow ($\omega$ = 40.43 rad/s), and (e) stable banded flow ($\omega$ = 77.75 rad/s).}
  \label{fig:setup2}
\end{figure}

Fig. 3 presents the flow map for $C_{g} = 40\mathrm{wt\%}$ as a function of angular velocity $\omega$ and normalized surfactant concentration $C_\mathrm{s} / C_\mathrm{cr}$. The critical angular velocity for each flow transition, denoted as $\omega_{\mathrm{cr}}$, was found to depend strongly on $C_\mathrm{s} / C_\mathrm{cr}$, which reflects the effect of interfacial tension. Specifically, the values of $\omega_\mathrm{cr}$ for the transitions from disturbed stratified flow (or transitional droplet flow) to unstable banded flow (TR I), and from unstable banded flow to stable banded flow (TR II), decreases with increasing $C_\mathrm{s} / C_\mathrm{cr}$. To generalize and rationalize the flow map, it is more appropriate to express it in terms of the Reynolds number \textit{Re} and Weber number \textit{We}, which are defined as follows:
\begin{equation}
    Re=\frac{\rho_\mathrm{m}R_\mathrm{i}\omega d}{\mu_{\mathrm{m}}}
\end{equation}
\begin{equation}
    We=\frac{\rho_\mathrm{m}d(R_\mathrm{i}\omega)^2}{\sigma}
\end{equation}
where, $\rho_\mathrm{m}\ \mathrm{[kg/m^3]}$ and $\mu_\mathrm{m}\ \mathrm{[Pa\cdot s]}$ are the mixture density and viscosity. According to Campero and Vigil (1999), they are defined as follows:
\begin{equation}
    \rho_\mathrm{m}=\rho_\mathrm{c}(1-\varepsilon)+\rho_{\mathrm{d}}\varepsilon
\end{equation}
\begin{equation}
    \mu_\mathrm{m}=\mu_{\mathrm{c}}^{(1-\varepsilon)}\mu_\mathrm{d}^\varepsilon
\end{equation}

\begin{figure}[htbp]
  \centering
  \includegraphics[width=1\linewidth]{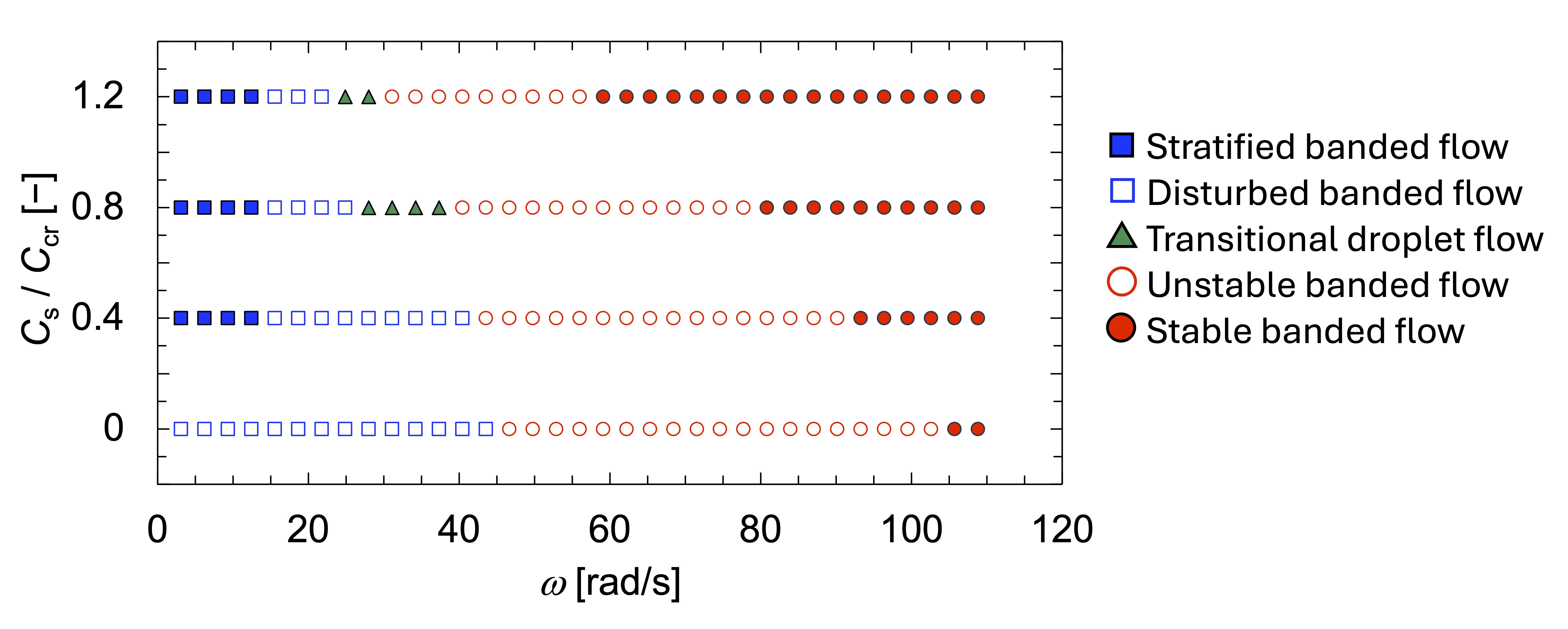}
  \caption{Flow map for $C_\mathrm{g}$ = 40wt\% as a function of angular velocity $\omega$ and normalized surfactant concentration $C_\mathrm{s} / C_\mathrm{cr}$.}
  \label{fig:setup3}
\end{figure}

Flow maps plotted as a function of \textit{We} and \textit{Re} are presented in Fig. 4 for each ${C}_{\mathrm{g}}$. Approximate boundary lines corresponding to TR I and TR II are also shown in Fig. 4. These transition boundaries can be approximated by a power-law relationship of the form $\textit{We}\sim \textit{Re}^{\textit{A}}$, where \textit{A} is a fitting exponent. The exponents \textit{A} for transition boundaries TR I and TR II are summarized in Table 3. In both cases, the exponent \textit{A} fallen within the range 0 < \textit{A} < 1. When \textit{A} = 0, the critical \textit{We} at the transition is independent of \textit{Re}, and thus remains constant. In contrast, when \textit{A} = 1, the critical condition implies that $We/Re$ is constant, which corresponds to a constant Capillary number (\textit{Ca}). A constant critical \textit{We} indicates that the transition is governed primarily by the balance between inertial forces and interfacial tension. In contrast, a constant \textit{Ca} implies that the transition is dominated by the competition between viscous forces and interfacial tension. In TR I, the exponent \textit{A} varied only slightly with $C_\mathrm{g}$, consistently falling within the range of 0.40–0.57. This suggests that the mechanism underlying TR I is analogous across different $C_\mathrm{g}$ systems, even though the absolute values of \textit{Re} and \textit{We} at the transition point vary. In addition, since the exponent \textit{A} in TR I falls within the range of approximately 0.40 to 0.57, the transition does not occur under a condition of constant \textit{We} or \textit{Ca}. This suggests that the transition TR I is governed by a combined effect of inertial forces, viscous forces, and interfacial tension.
For transition II, at $C_\mathrm{g} = 0\ \mathrm{and}\ 20\mathrm{wt}\%$, the exponent \textit{A} remained within a range similar to that for TR I, i.e., \textit{A} = 0.57 $(C_{\mathrm{g}} = 0\mathrm{wt}\%)$ and 0.49 $(C_{\mathrm{g}} = 20\mathrm{wt}\%)$. Hence, at relatively low $C_\mathrm{g}$, transition TR II is also considered to be governed by the combined effect of inertial forces, viscous forces, and interfacial tension, as in TR I. However, at higher $C_\mathrm{g}$ values of 40 and 60wt\%, the value of \textit{A} decreased; in particular, it became quite small ($\textit{A} \approx 0$) at $C_\mathrm{g} = 60\mathrm{wt}\%$. This indicates that, at $C_\mathrm{g} = 60\mathrm{wt}\%$, the critical Weber number for transition TR II is independent of \textit{Re}. This behavior suggests that, although increasing $C_\mathrm{g}$ raises the continuous-phase viscosity and therefore the shear stress acting on droplets (for a given velocity gradient), the onset of breakup at high $C_\mathrm{g}$ is not governed by viscous forces. Instead, the breakup transition is determined by the balance between inertial forces and interfacial tension, as indicated by the constant critical \textit{We}. In this regime, breakup occurs only when inertial forces overcome interfacial tension.
In addition, the viscosity ratio between the dispersed and continuous phases is known to influence droplet breakup behavior. Komrakova et al. (2015) reported that droplets with a higher viscosity than the surrounding continuous phase tend to elongate more and are less prone to breakup. In our system, as indicated in Table 1, the dispersed-phase viscosity was significantly higher than that of the continuous phase at low $C_\mathrm{g}$, resulting in a relatively large viscosity ratio. As $C_\mathrm{g}$ increased, the viscosity of the continuous phase increased, leading to a lower viscosity ratio and, consequently, droplets that were more susceptible to breakup as the stabilizing influence of viscosity diminished. This trend is consistent with the observed decrease in the exponent \textit{A} for transition TR II at higher $C_\mathrm{g}$.

\begin{figure}[htbp]
  \centering
  \includegraphics[width=1\linewidth]{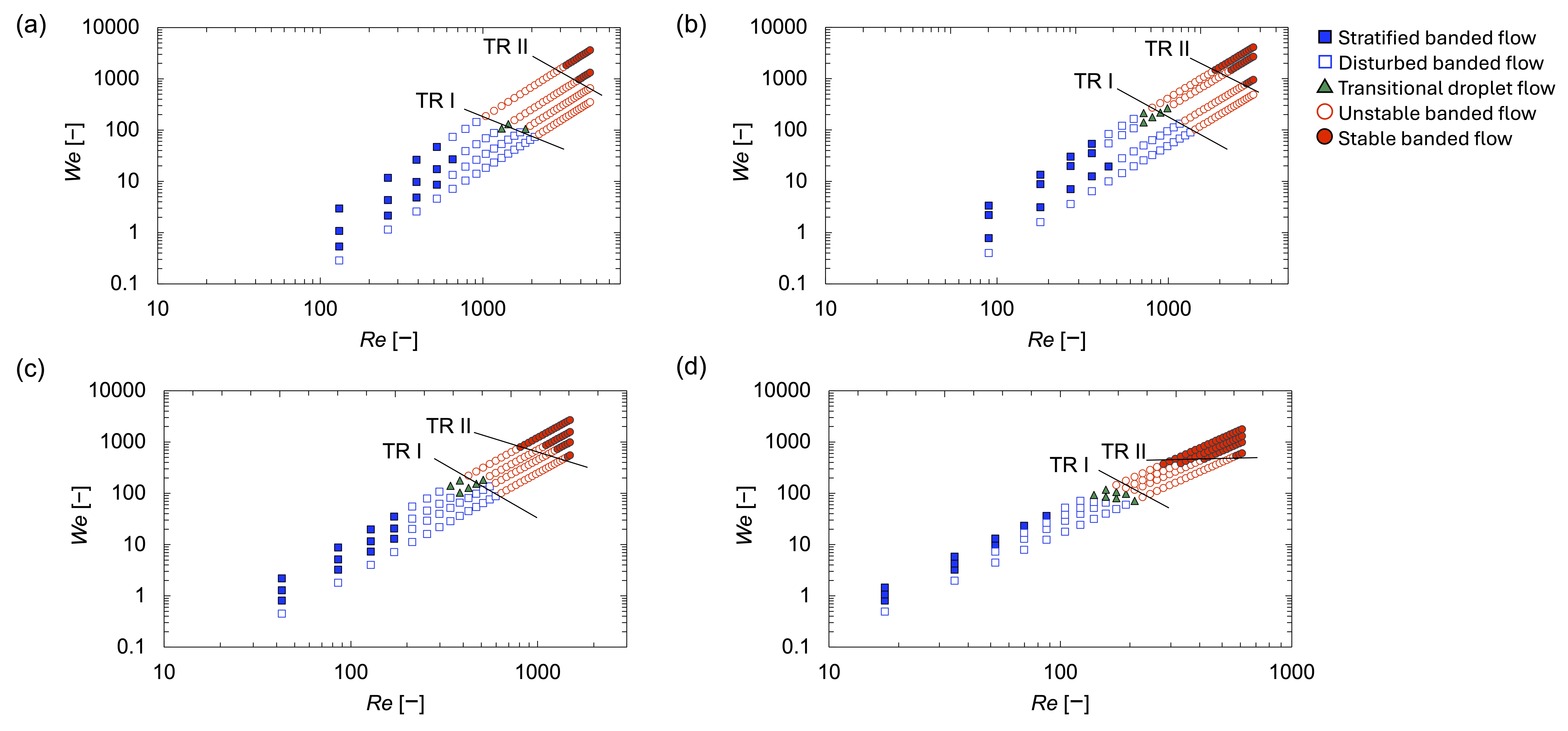}
  \caption{Flow map as a function of Reynolds number, \textit{Re}, and Weber number, \textit{We}: (a) $C_\mathrm{g}$ = 0wt\%, (b) $C_\mathrm{g}$ = 20wt\%, (c) $C_\mathrm{g}$ = 40wt\%, and (d) $C_\mathrm{g}$ = 60wt\%.}
  \label{fig:setup4}
\end{figure}

\begin{table}[htbp]
  \centering
  \caption{Power-law exponents $A$ for transition boundaries TR I and TR II as a function of continuous phase concentration $C_g$.}
  \label{tab:powerlaw}
  \begin{tabular}{c c c}
    \toprule
    $C_g$ [wt\%] & TR I & TR II \\
    \midrule
    0  & 0.40 & 0.57 \\
    20 & 0.55 & 0.49 \\
    40 & 0.57 & 0.30 \\
    60 & 0.54 & 0.03 \\
    \bottomrule
  \end{tabular}
\end{table}

In the stable banded flow regime, quite fine droplets are obtained, as shown in Fig. 2(e); hence, a high mass transfer rate in liquid–liquid chemical reactions or extraction is expected due to the large specific interfacial area. However, the mechanism of the transition from unstable to stable banded flow is still unclear. As mentioned in the introduction, Zhu and Vigil (2001) showed that the banded structure results from droplet migration from the vortex periphery to the vortex core by centrifugal effects. They introduced a non-dimensional parameter to estimate the transition. However, this parameter requires two quantities that can only be obtained from CFD simulation: the droplet azimuthal velocity relative to the vortex rotational axis and the turbulent viscosity. Their simulation also involved several simplified assumptions, including treating droplets as spheroidal particles. In addition, they used Haas’s equation (Haas, 1987) to estimate droplet sizes. Because this equation is valid only in high-Re turbulent flow regimes, it is not suitable for the present study with moderate Re values. Indeed, in our case Haas’s equation predicts unrealistically large droplets, even exceeding the gap width. Therefore, further consideration of the transition to stable banded flow is needed.
We hypothesize that the banded structure is formed by droplet trapping into the vortex core region due to their reduced inertia. Desmet et al. (1996) experimentally showed that a Taylor cell is divided into two parts: the vortex core and outer regions. According to Rudman (1998), if small particles are introduced, they are trapped in the core region, referred to as the “retention zone,” approximately for ${Re} \sim 400\text{\textendash}1000$. This range corresponds to the wavy vortex flow regime, which coincides with the conditions where stable banded flow is observed at $C_\mathrm{g} = 40\ \mathrm{and}\ 60\mathrm{wt}\%$. Rudman (1998) also showed that particles do not move into the outer region (the “chaotic flow region,” based on KAM theory) unless they have sufficient inertia. Thus, a similar trapping mechanism into the vortex core is assumed to cause the droplet banded structure.
To assess droplet inertia, the Stokes number, St, was calculated by assuming droplets to be spheroidal particles (Ouellette et al., 2008):
\begin{equation}
    St=\frac{1}{18}\frac{\rho_\mathrm{d}}{\rho_\mathrm{c}} \left( \frac{d_\mathrm{d}}{d_\mathrm{c}}\right)^2 Re
\end{equation}
where $d_\mathrm{d}$ is the average droplet diameter in the unstable and stable banded flow regimes. As mentioned in the experimental section, $d_\mathrm{d}$ was measured at the lower energy dissipation rate ($\varepsilon$ = 0.05) because at $\varepsilon$ = 0.1 droplet overlap prevented accurate measurements. Moreover, at lower $C_\mathrm{g}$ (= 0 and 20wt\%) and $C_\mathrm{s} / \textit{C}_\mathrm{cr} = 0$, accurate droplet sizes could not be measured due to their relatively large size even in the stable banded region. Thus, measurements were limited to $C_\mathrm{g}$ = 40 and 60wt\% and $C_\mathrm{s} / \textit{C}_\mathrm{cr}$ = 0.4, 0.8, and 1.2. Several dozen droplets in recorded images were analyzed using ImageJ software. The average droplet diameter dd as a function of \textit{We} is shown in Fig. 5. For each $C_\mathrm{g}$ condition, based on least-squares fitting, the diameter can be expressed as $d_\mathrm{d} = 55.65 We^{-0.55}$ for $C_\mathrm{g}$ = 40wt\% and $d_\mathrm{d} = 37.90 We^{-0.57}$ for $C_\mathrm{g}$ = 60wt\%. The fitted curves are also plotted in Fig. 5. Substituting the estimated droplet diameters into Eq. (5), \textit{St} was calculated. The relationship between \textit{St} and \textit{Re} in the unstable and stable banded flow regimes is shown in Fig. 6. Figure 6 indicates that the transition from unstable banded to stable banded flow (TR II) occurs when \textit{St} falls below a critical value. In particular, at $C_\mathrm{g}$ = 60wt\%, the critical \textit{St} is approximately 1, regardless of $C_\mathrm{s} / \textit{C}_\mathrm{cr}$. This means that stable banded flow results from the loss of droplet inertia with increasing \textit{Re}, owing to (i) droplet micronization by higher shear forces and (ii) enhancement of flow inertia. Combining these results (including flow visualization experiments) with Rudman’s (1998) theory, it is inferred that stable banded flow arises from droplet trapping into the vortex core. Further investigation under wider conditions (e.g., axial \textit{Re} and different combinations of continuous and dispersed fluids) will be required for a more complete understanding of the banded flow mechanism.

\begin{figure}[htbp]
  \centering
  \includegraphics[width=0.7\linewidth]{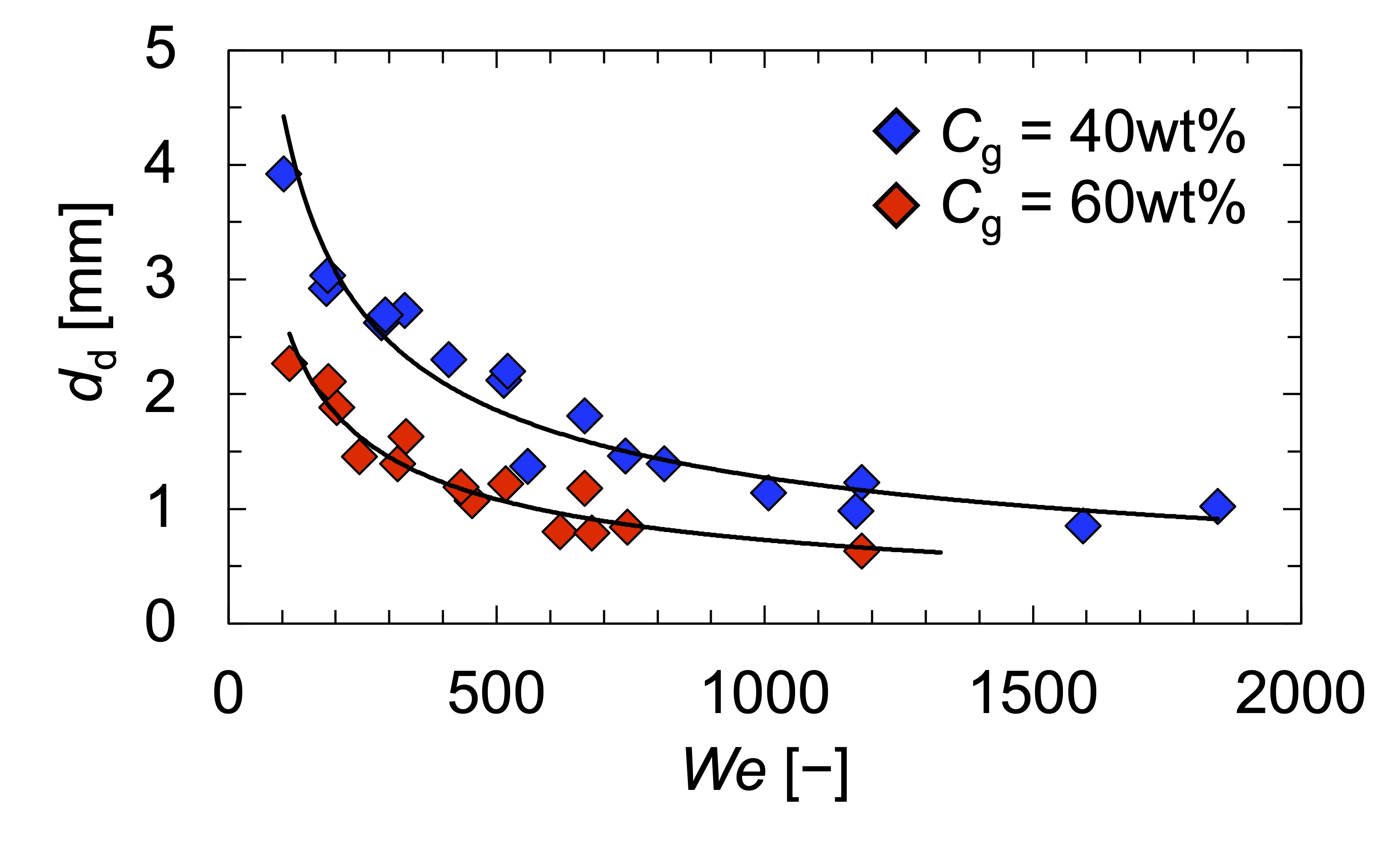}
  \caption{Average droplet diameter $d_\mathrm{d}$ as a function of Weber number, \textit{We}.}
  \label{fig:setup5}
\end{figure}

\begin{figure}[htbp]
  \centering
  \includegraphics[width=1\linewidth]{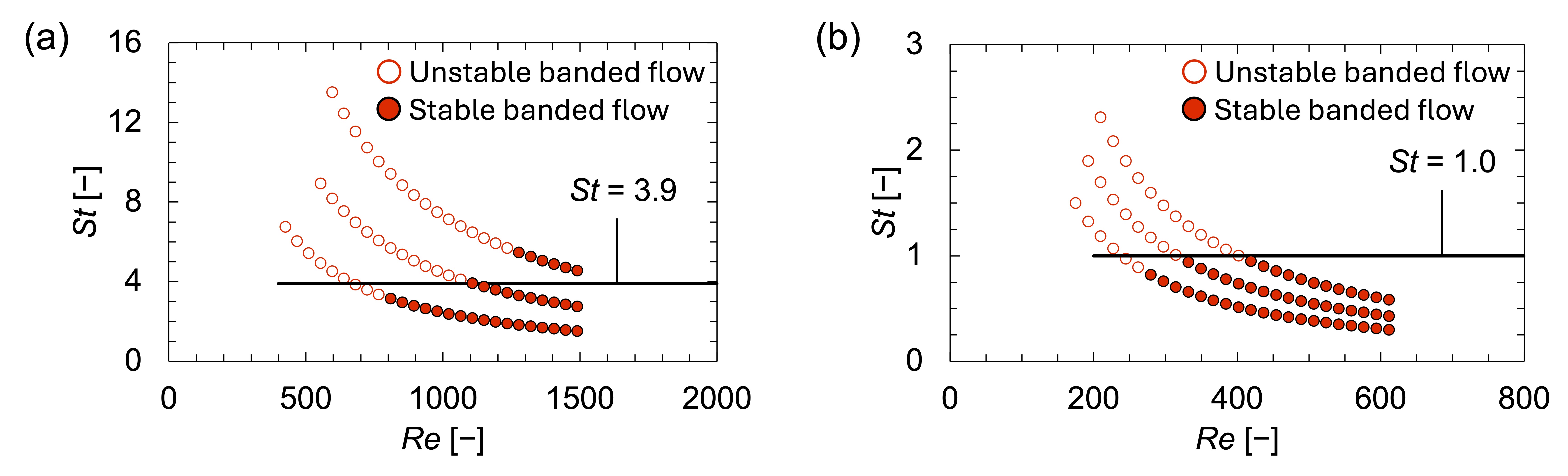}
  \caption{Relationship between \textit{St} and \textit{Re} in the unstable and stable banded flow regimes: (a) $C_\mathrm{g}$ = 40wt\% and (b) $C_\mathrm{g}$ = 60wt\%.}
  \label{fig:setup6}
\end{figure}

\section{Conclusions}
This study experimentally investigated the flow transition of liquid–liquid two-phase Taylor–Couette flow with increasing circumferential Reynolds number, Re. In particular, the study focused on the effect of continuous-phase viscosity by employing glycerol–water solutions at various concentrations as the continuous phase, with soybean oil as the dispersed phase. Flow patterns were recorded with a high-resolution video camera while gradually increasing the rotational speed of the inner cylinder.

From the recorded videos, the flow was classified into five regimes: stratified flow, disturbed stratified flow, transitional droplet flow (observed only in limited cases), unstable banded flow, and stable banded flow. A flow map was constructed using two dimensionless numbers, \textit{Re} and Weber number (\textit{We}). Both critical conditions—from disturbed stratified (or transitional droplet) to unstable banded flow (TR I), and from unstable to stable banded flow (TR II)—were found to be approximately expressed by a power-law relation, $We \sim Re^A$, where \textit{A} is a fitting exponent. For TR I, \textit{A} showed no clear dependence on continuous-phase viscosity, ranging between 0.40 and 0.57, suggesting that the transition is governed by a combined effect of inertial forces, viscous forces, and interfacial tension. In contrast, for TR II, \textit{A} decreased at higher $C_\mathrm{g}$ values of 40 and 60wt\%, and in particular became small ($\textit{A} \approx 0$) at $C_\mathrm{g}$ = 60wt\%. This indicates that, although increasing $C_\mathrm{g}$ raises the continuous-phase viscosity and hence the shear stress acting on droplets (for a given velocity gradient), the onset of breakup at high Cg is not governed by viscous forces.

To further examine the mechanism of stable banded flow, droplet inertia was evaluated using the Stokes number, \textit{St}, with the average droplet diameter obtained from image analysis. It was found that stable banded flow occurred when \textit{St} fell below a critical value. In particular, for the highest $C_\mathrm{g}$ (= 60wt\%), where the banded structure was most clearly observed, the critical \textit{St} was approximately 1. These results suggest that the stable banded structure is formed by the loss of droplet inertia and subsequent trapping into the vortex core region.

\section{Acknowledgement}
This research was partially supported by JSPS KAKENHI grant numbers 21K14450, 21KK0261, and 25K08368.

\section{References}  

Aksamija, E., Weinländer, C., Sarzio, R., Siebenhofer, M., 2015. The Taylor-Couette disc contactor: a novel apparatus for liquid/liquid extraction. Sep. Sci. Technol, 50 (18), 2844–2852.

Baier, G., Graham, M. D., Lightfoot, E. N., 2000. Mass transport in a novel two-fluid taylor vortex extractor. AIChE J. 46 (12), 2395–2407.

Campbell, C., Olsen, M. G., Vigil, R. D., 2019. Flow regimes in two-phase hexane/water semibatch vertical Taylor vortex flow. J. Fluid Eng. 141 (11), 111203.

Campbell, C., Olsen, M. G., Vigil, R. D., 2020. Droplet size distributions in liquid–liquid semi-batch Taylor vortex flow. AIP Adv. 10, 085316.

Campero, R. J., Vigil, R. D., 1997. Spatiotemporal patterns in liquid-liquid Taylor-Couette-Poiseuille flow. Phys. Rev. Lett. 79 (20), 3897–3900.

Campero, R. J., Vigil, R. D., 1999. Flow patterns in liquid-liquid Taylor–Couette–Poiseuille flow. Ind. Eng. Chem. Res. 38 (3), 1094–1098.

Davis, M. W., Weber, E. G., 1960. Liquid-liquid extraction between rotating concentric cylinders. Ind. Eng. Chem. 52 (11), 929–934.

Dandelia, H., Kant, R., Narayanan, V., 2002. Optimal control of growth of instabilities in Taylor–Couette flow. Phys. Fluids 34, 044106.

Desmet, G., Verelst, H., Baron, G. V., 1996. Local and global dispersion effects in Couette-Taylor flow—I. Description and modeling of the dispersion effects. Chem. Eng. Sci. 51 (8), 1287–1298.

Dluska, E., Markowska-Radomska, A., 2010. Regimes of multiple emulsions of W1/O/W2 and O1/W/O2 type in the continuous Couette-Taylor flow contactor. Chem. Eng. Technol. 33 (1), 113–120.

Grafschafter, A., Aksamija, E., Siebenhofer, M., 2016. The Taylor-Couette disc contactor. Chem. Eng. Technol. 39 (11), 2087–2095.

Haas, P. A., 1987. Turbulent dispersion of aqueous drops in organic liquids. AIChE J. 33 (6), 987–995.

Hori, N., Ng, C. S., Lohse, D., Verzicco, R., 2023. Interfacial-dominated torque response in liquid–liquid Taylor–Couette flows. J. Fluid Mech. 956, A15.

Hubacz, R., Wroński, S., 2004. Horizontal Couette–Taylor flow in a two-phase gas–liquid system: flow patterns. Exp. Therm. Fluid Sci. 28, 457–466.

Joseph, D. D., Nguyen, K., Beavers, G. S., 1984. Non-uniqueness and stability of the configuration of flow of immiscible fluids with different viscosities. J. Fluid Mech. 141, 319–345.

Joseph, D. D., Renardy, Y., Renardy, M., Nguyen, K., 1985. Stability of rigid motions and rollers in bicomponent flows of immiscible liquids. J. Fluid Mech. 153, 151–165.

Komrakova, A. E., Shardt, O., Eskin, D., Derksen, J. J., 2015. Effects of dispersed phase viscosity on drop deformation and breakup in inertial shear flow. Chem. Eng. Sci. 126, 150–159.

Kusumastuti, A., Najibulloh, G. M., Qudus, N., Anis, S., 2019. Flow regime analysis of Taylor-Couette column for emulsion liquid membrane applications. AIP Conf. Proc. 2124, 020047.

Li, B., Li, G., Yang, X., Yao, K., Guo, Y., 2023. Effect of inner cylinder configuration on liquid droplet dispersion and size distributions in a Taylor-Couette reactor. Chem. Eng. Sci. 267, 118362.

Masuda, H., Horie, T., Hubacz, R., Ohmura, N., 2013. Process intensification of continuous starch hydrolysis with a Couette–Taylor flow reactor. Chem. Eng. Res. Des. 91 (11), 2259–2264.

Masuda, H., Shimoyamada, M., Ohmura, N., 2019a. Heat transfer characteristics of Taylor vortex flow with shear-thinning fluids. Int. J. Heat Mass Transf. 130 (9), 274–281.

Masuda, H., Hubacz, R., Shimoyamada, M., Ohmura, N., 2019b. Numerical simulation of sterilization processes for shear-thinning food in Taylor-Couette flow systems. Chem. Eng. Technol. 42 (4), 859–866.

Matsumoto, M., Masuda, H., Hubacz, R., Horie, T., Iyota, H., Shimoyamada, M., Ohmura, N., 2021. Enzymatic starch hydrolysis performance of Taylor-Couette flow reactor with ribbed inner cylinder. Chem. Eng. Sci. 231, 116270.

Ouellette, N. T., O’Malley, P. J. J., Gollub, J. P., 2008. Transport of finite-sized particles in chaotic flow. Phys. Rev. Lett. 101, 174504.

Renardy, Y., Joseph, D. D., 1985. Couette flow of two fluids between concentric cylinders. J. Fluid Mech. 150, 381–394.

Rudman, M., 1998. Mixing and particle dispersion in the wavy vortex regime of Taylor–Couette flow. AIChE J. 44 (5), 1015–1026.

Sathe, M. J., Deshmukh, S. S., Joshi, J. B., Koganti, S. B., 2010. Computational fluid dynamics simulation and experimental investigation: Study of two-phase liquid–liquid flow in a vertical Taylor–Couette contactor. Ind. Eng. Chem. Res. 49 (1), 14–28.

Wang, B., Tao, S., 2022. Synthesis of micro-/nanohydroxyapatite assisted by the Taylor–Couette flow reactor. ACS Omega 7, 44057–44064.

Wen, J., Zhang, W. -Y., Ren, L. -Z., Bao, L. -Y., Dini, D., Xi, H. -D., Hu, H. -B., 2020. Controlling the number of vortices and torque in Taylor–Couette flow. J. Fluid Mech., 901, A30.

Zeng, Z., Wang, Z., Su, Y., Yao, C., Chen, G., 2024. Liquid–liquid flow characteristics and mass transfer enhancement in a Taylor–Couette microreactor. AIChE J., 70 (10), e18528.

Zhang, B., Coquerel, G., Park, B. J., Kim, W. -S., 2024. Chiral symmetry breaking of sodium chlorate in a Taylor vortex flow. Cryst. Growth Des. 24, 1042–1050.

Zhu, X., Vigil, R. D., 2001. Banded liquid–liquid Taylor-Couette-Poiseuille flow. AIChE J. 47 (9), 1932–1940.


\end{document}